\documentclass[10pt,conference]{IEEEtran}
\IEEEoverridecommandlockouts

\usepackage{xcolor,soul,framed} 
\usepackage{algorithmic}
\usepackage{algorithm}
\colorlet{shadecolor}{yellow}
\usepackage[pdftex]{graphicx}
\DeclareGraphicsExtensions{.pdf,.jpeg,.png}
\usepackage[cmex10]{amsmath}
\usepackage{amssymb}
\usepackage{array}
\usepackage{mdwmath}
\usepackage{mdwtab}
\usepackage{eqparbox}
\usepackage{url}
\usepackage{bm}
\usepackage{pifont}
\usepackage{multirow}  
\usepackage{multicol}  
\usepackage{makecell}
\usepackage {cite}
\usepackage{booktabs}
\usepackage{multirow}
\usepackage{threeparttable}
\newcommand{\Min}{\operatorname{Minimize}}

\renewcommand{\baselinestretch}{0.94}

\begin{document}
\bstctlcite{IEEEexample:BSTcontrol}
    \title{PLS-Assisted Offloading for \\Edge Computing-Enabled Post-Quantum Security \\in Resource-Constrained Devices


}

   \author{
    \IEEEauthorblockN{Hamid Amiriara, Mahtab Mirmohseni, and Rahim Tafazolli}
    \IEEEauthorblockA{
        \textit{Institute for Communication Systems, University of Surrey, Guildford, U.K.}\\
        Email: \{h.amiriara, m.mirmohseni, r.tafazolli\}@surrey.ac.uk \vspace{-.9em}
    }
}

\markboth{IEEE TRANSACTIONS ON COMMUNICATIONS, VOL.~1, NO.~1, March~2025} 
         {Hamid \MakeLowercase{\textit{et al.}}: Edge Computing-Enabled PQC Offloading Optimization}
\maketitle

\begin{abstract}
With the advent of post-quantum cryptography (PQC) standards, it has become imperative for resource-constrained devices (RCDs) in the Internet of Things (IoT) to adopt these quantum-resistant protocols. However, the high computational overhead and the large key sizes associated with PQC make direct deployment on such devices impractical. To address this challenge, we propose an edge computing-enabled PQC framework that leverages a physical-layer security (PLS)-assisted offloading strategy, allowing devices to either offload intensive cryptographic tasks to a post-quantum edge server (PQES) or perform them locally.
Furthermore, to ensure data confidentiality within the edge domain, our framework integrates two PLS techniques: offloading RCDs employ wiretap coding to secure data transmission, while non-offloading RCDs serve as friendly jammers by broadcasting artificial noise to disrupt potential eavesdroppers.
Accordingly, we co-design the computation offloading and PLS strategy by jointly optimizing the device transmit power, PQES computation resource allocation, and offloading decisions to minimize overall latency under resource constraints. Numerical results demonstrate significant latency reductions compared to baseline schemes, confirming the scalability and efficiency of our approach for secure PQC operations in IoT networks.
\end{abstract}



\section{Introduction} \label{Introduction} \vspace{-.1cm}
The Internet of Things (IoT) extends Internet connectivity to numerous embedded devices (e.g., sensors and actuators) and has seen rapid adoption in areas such as smart transportation, home automation, and e-healthcare \cite{Elmehdi2024}. Despite its advantages, securing these resource-constrained devices (RCDs) remains challenging due to their limited computing power, memory, and battery life. Such constraints complicate the deployment of modern cryptographic schemes, as these schemes often demand substantial computational resources \cite{patrick2016}.

Conventional cryptographic algorithms, such as Rivest–Shamir–Adleman (RSA) and elliptic curve cryptography (ECC), have secured digital systems for decades. However, quantum computing poses a critical threat as algorithms such as Shor’s can efficiently break these schemes \cite{shor1994}.
To counter quantum threats, the National Institute of Standards and Technology (NIST) is standardizing post-quantum cryptography (PQC) algorithms, including Kyber, Dilithium, Falcon, and SPHINCS\textsuperscript{+} \cite{NIST_PQC, NIST_ThreeDraft}. However, these algorithms introduce new challenges for RCDs due to their high computational demands, large key sizes, and significant energy costs. To better illustrate these challenges, Table~\ref{tab:comparison} compares these schemes to classical cryptography in terms of key sizes, ciphertext/signature sizes, and computational overhead. As shown in Table~\ref{tab:comparison}, even at its lowest security level (Dilithium-2), Dilithium requires a 1312-byte public key and a 2420-byte signature, approximately five and nine times larger than RSA, respectively. Meanwhile, SPHINCS\textsuperscript{+} has compact keys, but uses hash-based signing operations with extremely high CPU cycles per bit, and Falcon requires floating-point operations unsuitable for many IoT platforms.

\begin{table}[b] 
\vspace{-.4cm}
\caption{Comparison of Key Sizes, Ciphertext (Ct)/Signature (Sig) Sizes (in bytes), and CPU-Cycle Overhead for Conventional Cryptographic and PQC Algorithms.}
\vspace{-.3cm}
\label{tab:comparison}
\centering
\begin{tabular}{lp{.9cm}p{.7cm}p{.7cm}p{0.8cm}l}
\hline
\textbf{Algorithm} & \textbf{Security level} & \textbf{Pub. key} & \textbf{Priv. key} & \textbf{Ct/Sig} & \makecell{\textbf{CPU*}\\\textbf{cycles/bit}}  \\ \hline 
RSA-2048          & NA          & 256                      & 256                        & 256                      & 113                    \\
ECC-256           & NA          & 64                       & 32                         & 256                      & 281                    \\
Kyber-512         &1             & 800                      & 1,632                      & 768                      & 2,193              \\
Kyber-768         &3             & 1,184                    & 2,400                      & 1,088                    & 3,577            \\
Kyber-1024        &5             & 1,568                    & 3,264                      & 1,568                    & 5,499            \\
Dilithium-2       &1             & 1,312                    & 2,528                      & 2,420                    & 24,051            \\
Dilithium-3       &3             & 1,952                    & 4,000                      & 3,293                    & 36,287            \\
Dilithium-5       &5             & 2,592                    & 4,864                      & 4,595                    & 33,085            \\
Falcon-512        &1             & 897                      & 1,281                      & 690                      & 148,791           \\
Falcon-1024       &3             & 1,793                    & 2,305                      & 1,330                    & 326,105           \\
SPHINCS\textsuperscript{+}-128f     &1             & 32                       & 64                         & 17,088                   & 2,038,919          \\
SPHINCS\textsuperscript{+}-192f     &3             & 48                       & 96                         & 35,664                   & 2,686,303          \\
SPHINCS\textsuperscript{+}-256f     &5             & 64                       & 128                        & 49,856                   & 6,070,970        \\ \hline 
\end{tabular}
\vspace{-.05cm}
\begin{tablenotes}
\footnotesize
\item * CPU cycles/bit measured on ARM Cortex-M4 for encryption or signing of a 256-bit message, averaged over multiple runs \cite{Kannwischer2019, C_k_Kyber, C_k_RSA_WolfSSLBenchmarks}.
\end{tablenotes}\vspace{-.2cm}
\end{table}
A promising strategy for addressing these challenges is to offload computations to more capable edge nodes, thereby alleviating on-device overhead \cite{ehealth2024,edge3}. While offloading PQC-related tasks can reduce resource consumption and enhance performance, transmitting sensitive data over wireless links exposes RCDs to eavesdroppers (EVEs). Conventional encryption-based approaches may still be susceptible to quantum attacks \cite{Zhang2018}, and often require an RCD to encrypt data, which then must be decrypted and re-encrypted at the edge- greatly increasing computational complexity and latency.

Alternatively, physical-layer security (PLS) techniques, including wiretap coding and friendly jamming, can provide information-theoretic security directly over the wireless medium \cite{Wyner1975, Hamid}. Despite this potential, integrating PLS into an offloading framework requires a co-design approach that addresses both RCD constraints and limited network resources. 

In this paper, we propose a PLS-assisted offloading scheme tailored to PQC in IoT networks. To reduce device-side overhead, the proposed framework allows each RCD to either process cryptographic tasks locally (as a non-offloading RCD) or offload them to a post-quantum edge server (PQES) (as an offloading RCD). To protect offloaded data, we integrate two PLS techniques: Offloading RCDs use wiretap coding for secure transmission, while non-offloading RCDs act as friendly jammers by broadcasting artificial noise to degrade EVE's reception. Accordingly, we formulate a joint computation offloading and PLS co-design problem to minimize total latency by jointly optimizing transmit power (for offloading or jamming), computation resource allocation at the PQES, and the offloading decisions under transmit-power and computing-resource constraints.

The main contributions of this paper are outlined as follows:\begin{itemize} 
\item We introduce a PLS-assisted offloading framework for PQC in IoT that not only alleviates the computational load on RCDs but also secures offloaded data.
\item We formulate a mixed-integer non-convex optimization problem aimed at minimizing the total latency across all RCDs by jointly optimizing RCD transmit power (for offloading/jamming), PQES computation resource allocation, and the offloading mode decisions. To solve this complex problem, we employ a variable separation and alternating optimization (AO) approach combined with successive convex approximation (SCA).
\item To capture practical PQC-induced latency, our framework assigns CPU cycles per bit from Table~\ref{tab:comparison}, accounting for varying PQC algorithms and security levels across RCDs.
\item Through extensive simulations, we demonstrate that our proposed framework achieves significant latency reduction while maintaining robust security in PQC-enabled IoT deployments.
\end{itemize}

\section{System Model and Problem Formulation} \label{System_Model} \vspace{-.1cm}
\subsection {System Model}
\begin{figure}[!t]
  \begin{center}
  \includegraphics[width=7.3cm]{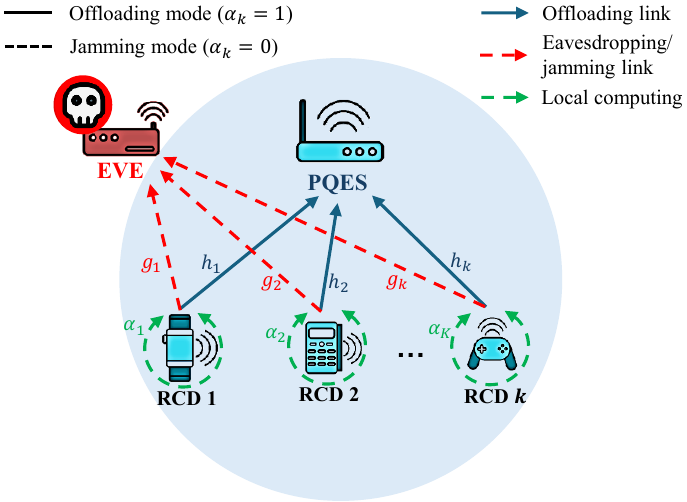}\\
  \vspace{-0.3cm}
  \caption{PLS-assisted offloading in edge computing-enabled PQC systems.}\label{fig1}
  \end{center}
   \vspace{-.8cm}
\end{figure}

\vspace{-.1cm}
As illustrated in Fig.~\ref{fig1}, we consider an edge computing-enabled PQC system comprising a single PQES and $K$ RCDs, under the threat of a malicious EVE. 

\noindent\textbf{Offloading and Jamming Strategy:}
Each RCD can operate in one of two modes: (i) As an \textit{offloading RCD}, it offloads computationally intensive PQC tasks to the PQES. (ii) As a \textit{non-offloading/jamming RCD}, it processes these tasks locally while simultaneously acting as a friendly jammer by broadcasting artificial noise. The mode selection for the $k$-th RCD is indicated by a binary variable $\alpha_k$, where $\alpha_k = 1$ represents an offloading RCD and $\alpha_k = 0$ denotes a non-offloading RCD. Accordingly, the set of offloading RCDs is $\mathcal{K}_\text{Off} = \{ k \mid \alpha_k = 1,\ k \in \mathcal{K} \}$, and the set of non-offloading RCDs is $\mathcal{K}_\text{Non-off} = \{ k \mid \alpha_k = 0,\ k \in \mathcal{K} \}$, where $\mathcal{K} \triangleq \{1,2,\dots, K\}$. 

In this framework, the PQES allocates computing resources to efficiently process cryptographic tasks offloaded by devices in $\mathcal{K}_\text{Off}$, while devices in $\mathcal{K}_\text{Non-off}$ perform PQC locally and enhance security by transmitting artificial noise to jam potential EVE. Specifically, offloading RCDs use their transmit power $p_k, k \in \mathcal{K}_\text{Off}$ to send tasks to the PQES, whereas non-offloading RCDs use $p_k,k \in \mathcal{K}_\text{Non-off}$ to transmit jamming signals. This co-designed approach ensures that all RCDs contribute to both computational efficiency and data confidentiality.

\noindent\textbf{Edge Computing Offloading Model:} Let $d_k$ denote the data size (in bits) of the data that must undergo cryptographic operation for the $k$-th RCD. Furthermore, let $c_k$ represent the number of CPU cycles required per bit of data, which is determined by the specific PQC algorithm and the security level required by the $k$-th RCD (see Column 6 of Table~\ref{tab:comparison}).

\noindent \textit{1) Local Computing:} Define $f_0$ as the local computing capacity (in CPU cycles per second) of each RCD. If the $k$-th RCD performs cryptographic processing locally, the corresponding local computing latency is given by \cite{Dai2018},
\vspace{-0.1cm} \begin{equation}
T_k^{\text{Loc}} = \frac{d_k c_k}{f_0}, \quad \forall k \in \mathcal{K}_\text{Non-off}.
\label{eq:local_latency}
\vspace{-0.1cm} \end{equation}
\textit{2) Offloading to PQES:} 
For RCDs opting to offload, secure communication is enforced via wiretap coding and friendly jamming. The offloading rate for each RCD is defined as its achievable secrecy rate to the PQES, denoted as ${S\!R}_k$, ensuring the confidentiality of the data. Consequently, the transmission latency for the $k$-th RCD is expressed as,
\vspace{-0.1cm} \begin{equation}
T_k^{\text{Tra}} = \frac{d_k}{B \, {S\!R}_k}, \quad \forall k \in \mathcal{K}_\text{Off},
\label{eq:transmission_latency}
\vspace{-0.1cm} \end{equation}
where $B$ is the available communication bandwidth. Let $f_k$ represent the computational capacity allocated by the PQES to the $k$-th RCD. The computing latency at the PQES is then given by:
\vspace{-0.1cm} \begin{equation}
T_k^{\text{Off}} = \frac{d_k c_k}{f_k}, \quad \forall k \in \mathcal{K}_\text{Off}.
\label{eq:offloading_latency}
\vspace{-0.1cm} \end{equation}
\textit{3) Total Latency:} The overall latency for the $k$-th RCD's cryptographic task is thus:
\vspace{-0.1cm} \begin{equation}
T_k^{\text{Tot}} = (1-\alpha_k) \, T_k^{\text{Loc}} + \alpha_k \, (T_k^{\text{Tra}} + T_k^{\text{Off}}), \quad \forall k \in \mathcal{K}.
\label{eq:total_latency}
\vspace{-0.1cm} \end{equation}

\noindent\textbf{Secure Communication Model:}
We denote $h_k$ and $g_k$ as the respective channel gains from the $k$-th RCD to the PQES and to EVE. We assume that the PQES has perfect knowledge of the RCD's channel state information (CSI), i.e., $h_k$, and their computational requirements. However, the PQES has only partial knowledge of EVE's CSI, denoted by $g_k$. Following the standard practice in PLS literature \cite{uncertaintymodel1}, we adopt a deterministic uncertainty model for EVE's CSI given by $g_k = \tilde{g}_k + \Delta g_k$, where $\tilde{g}_k$ is the estimated channel gain for EVE, $\Delta g_k$ is the estimation error, and $|\Delta g_k| \leq \epsilon$, with $\epsilon \geq 0$ representing the maximum possible estimation error.

In PLS, the worst-case secrecy rate (bits/sec/Hz) for the $k$-th RCD under a wiretap coding strategy is:
\vspace{-0.1cm} \begin{equation}
{S\!R}_k = \big[ R_k^{\text{RCD}} - \max_{|\Delta g_k| \leq \epsilon} R_k^{\text{EVE}} \big]^+, \quad \forall k \in \mathcal{K}_\text{Off},
\label{eq:secrecy_rate}
\vspace{-0.1cm} \end{equation}
where $[x]^+ \triangleq \max\{0, x\}$. The terms $R_k^{\text{RCD}}$ and $R_k^{\text{EVE}}$ represent the data transmission rate for the $k$-th RCD to the PQES and the achievable data rate for EVE to wiretap the $k$-th RCD’s information, respectively, defined as
\vspace{-0.1cm} \begin{equation}
R_k^{\text{RCD}} = \log_2( 1 + \frac{p_k h_k}{\sum_{\substack{i \in \mathcal{K}, j \neq k}} p_j h_j + \sigma_\text{RCD}^2}),
\label{eq:RCD_rate}
\vspace{-0.1cm} \end{equation}
\vspace{-0.01cm} \begin{equation}
R_k^{\text{EVE}} = \log_2( 1 + \frac{p_k g_k}{\sum_{\substack{j \in \mathcal{K}, j \neq k}} p_j g_j + \sigma_\text{EVE}^2} ),
\label{eq:EVE_rate}
\vspace{-0.1cm} \end{equation}
with $\sigma_\text{RCD}^2$ and $\sigma_\text{EVE}^2$ denote the variance of the additive white Gaussian noise at the RCD and EVE, respectively.
\vspace{-0.1cm}
\subsection {Problem Formulation} \vspace{-.1cm}
Under this setup, we aim to minimize overall RCD latency by jointly optimizing the transmit power vector $\mathbf{p} \triangleq \{p_k,\, k \in \mathcal{K}\}$, computing capacity allocation vector $\mathbf{f} \triangleq \{f_k,\, k \in \mathcal{K}_\text{Off}\}$, and the offloading decisions vector $\boldsymbol{\alpha} \triangleq \{\alpha_k,\, k \in \mathcal{K}\}$. Thus, the joint optimization problem is formulated as
\begin{subequations}
\vspace{-0.1cm} \begin{align}
(\mathcal{P}):\quad \underset{\mathbf{p},\mathbf{f},\boldsymbol{\alpha}} \Min \quad & \sum_{k \in \mathcal{K}} T_k^{\text{Tot}} \label{eq:P_obj} \\[1mm]
\text{s.t.}\quad & \sum_{k \in \mathcal{K}} \alpha_k f_k \leq f^\text{Tot}, \label{eq:P_comp_cap} \\
& p_k \leq p^{\text{Max}}, \quad \ \forall k \in \mathcal{K}, \label{eq:power_constraint} \\
& \alpha_k \in \{0,1\}, \ \ \forall k \in \mathcal{K}, \label{eq:P_binary}
\vspace{-0.1cm} \end{align}
\end{subequations}
where $f^\text{Tot}$ is the total computing capacity of the PQES, and $p^{\text{Max}}$ is the maximum transmit power for each RCD.
Next, by introducing a vector of auxiliary variables $\mathbf{t}^\text{Th}\triangleq\{t^\text{Th}_k, k \in \mathcal{K}\}$, we can reformulate ($\mathcal{P}$) into an equivalent problem as
\begin{subequations}
\vspace{-0.1cm} \begin{align}
(\mathcal{P}.1):\quad \underset{\mathbf{p},\mathbf{f},\boldsymbol{\alpha}, \mathbf{t}^\text{Th}} \Min \quad & \sum_{k \in \mathcal{K}} t^\text{Th}_k \label{eq:P1_obj} \\[1mm]
\text{s.t.}\quad &T_k^{\text{Tot}} \leq t^\text{Th}_k,\quad \forall k \in \mathcal{K}, \label{eq:auxiliary}\\[1mm]
& \eqref{eq:P_comp_cap}, \eqref{eq:power_constraint}, \eqref{eq:P_binary},\nonumber
\vspace{-0.1cm} \end{align}
\end{subequations}
where $t^\text{Th}_k$ is the maximum allowable latency for $k$-th RCD.

Due to channel uncertainties that affect the secrecy rate in \eqref{eq:secrecy_rate}, constraint \eqref{eq:auxiliary} makes problem ($\mathcal{P}.1$) difficult to solve directly. Hence, to gain analytical tractability, we introduce a conservative upper bound on the latency terms, defined as:
\vspace{-0.1cm} \begin{equation}
T_k^{\text{Tot},\text{UB}} \triangleq (1-\alpha_k) \, T_k^{\text{Loc}} + \alpha_k ( \frac{d_k}{B \, {S\!R}_k^{\text{LB}}} + T_k^{\text{Off}}),
\label{eq:total_latency_upperbound}
\vspace{-0.2cm} \end{equation}
where ${S\!R}_k^{\text{LB}}$ is a conservative lower bound on the worst-case secrecy rate, defined as
\vspace{-0.1cm} \begin{equation}
{S\!R}_k^{\text{LB}} \triangleq \big[ R_k^{\text{RCD}} - {R}_k^{\text{EVE},\text{UB}} \big]^+, \quad \forall k \in \mathcal{K},
\label{eq:LSR_def}
\vspace{-0.2cm} \end{equation}
and
\vspace{-0.2cm} \begin{equation}
{R}_k^{\text{EVE},\text{UB}} = \log_2( 1 + \frac{p_k g_k^+}{\sum_{\substack{j \in \mathcal{K}, j \neq k}} p_j g_j^- + \sigma^2} ),
\label{eq:UR_EVE}
\vspace{-0.1cm} \end{equation}
where $g_k^+ = g_k + \epsilon$ and $g_j^- = g_j - \epsilon$ denote, respectively, the best- and worst-case channel gains for EVE from the PQES’s perspective.
With these modifications, problem ($\mathcal{P}.1$) is reformulated as:
\begin{subequations}
\vspace{-0.1cm} \begin{align}
(\mathcal{P}.2):\quad \underset{\mathbf{p},\mathbf{f},\boldsymbol{\alpha}, \mathbf{t}^\text{Th}} \Min \quad & \sum_{k \in \mathcal{K}} t^\text{Th}_k \label{eq:P2_obj} \\[1mm]
\text{s.t.}\quad & T_k^{\text{Tot},\text{UB}} \leq t_k^{\text{Th}}, \quad \forall k \in \mathcal{K}, \label{eq:P1_auxiliary} \\[1mm] 
& \eqref{eq:P_comp_cap}, \eqref{eq:power_constraint}, \eqref{eq:P_binary}.\nonumber
\vspace{-0.3cm} \end{align}
\end{subequations}
By enforcing $T_k^{\text{Tot},\text{UB}} \leq t_k^{\text{Th}}$ in \eqref{eq:P1_auxiliary}, we ensure that the actual latency $T_k^{\text{Tot}}$ also remains below $t_k^{\text{Th}}$ despite channel estimation uncertainties, thereby maintaining the robustness and validity of constraint \eqref{eq:auxiliary}.
\vspace{-.1cm} 
\section {Proposed Solution} \vspace{-.1cm} \label{Problem_Formulation}
Problem ($\mathcal{P}.2$) is inherently mixed-integer non-convex, making it challenging to find a globally optimal solution through standard methods. To address this, we propose a three-step AO algorithm. Specifically, we solve ($\mathcal{P}.2$) by alternately optimizing the RCDs’ transmit power $p_k$, computing capacity allocation $f_k$, and binary offloading decision variables $\alpha_k$.
\vspace{-0.1cm}
\subsection{Sub-problem 1: RCDs’ Transmit Power Optimization} \vspace{-.1cm} 
\label{Subproblem1}
With fixed computing capacity allocation $f_k$ and offloading decision vector $\alpha_k$, sub-problem 1 is defined as
\begin{subequations}
\vspace{-0.1cm} \begin{align}
(\mathcal{SP}1):\quad& \underset{\mathbf{p}, \mathbf{t}^\text{Th}} \Min \quad \sum_{k \in \mathcal{K}} t^\text{Th}_k \label{eq:SP1_obj} \\[1mm]
\text{s.t.} \quad & {S\!R}_k^{\text{LB}} \geq \frac{d_k}{B(t_k^{\text{Th}}-T_k^{\text{Off}})}, \quad \forall k \in \mathcal{K}_\text{Off}, \label{eq:SP1_cons}\\
& \eqref{eq:power_constraint}. \nonumber
\vspace{-0.4cm} \end{align}
\end{subequations}
Due to the non-concavity of $R_k^{\text{RCD}}$ and non-convexity of ${R}_k^{\text{EVE},\text{UB}}$, ($\mathcal{SP}1$) is non-convex. To address this issue, first, we express the RCD rate $R_k^{\text{RCD}}$ as the difference of two concave functions with respect to the transmit power vector $\textbf{p}$:
\vspace{-0.01cm} \begin{equation}
R_k^{\text{RCD}} = I(\textbf{p}) - J_k(\textbf{p}),
\label{eq:RCD_decomp}
\vspace{-0.01cm} \end{equation}
where $I(\textbf{p}) \triangleq \log_2\big(\sum_{k \in \mathcal{K}} p_k h_k + \sigma^2\big)$, and $J_k(\textbf{p}) \triangleq \log_2\big(\sum_{\substack{j \in \mathcal{K}, j \neq k}} p_j h_j + \sigma^2\big)$.
The non-concavity of $R_k^{\text{RCD}}$ arises from $J_k(\textbf{p})$. To handle this, we employ the SCA approach. Specifically, in the $r$-th iteration of the SCA algorithm, $J_k(\textbf{p})$ can be upper-bounded by:
\vspace{-0.1cm} \begin{equation}
\hat{J}_k^{(r)}(\textbf{p}) \triangleq J_k\big(\textbf{p}^{(r)}\big) + \frac{\sum_{\substack{i \in \mathcal{K}, i \neq k}} (p_i - p_i^{(r)}) h_i}{\ln 2 \big(\sum_{\substack{i \in \mathcal{K}, i \neq k}} p_i^{(r)} h_i + \sigma^2\big)},
\label{eq:J_taylor}
\vspace{-0.1cm} \end{equation}
which is the first-order Taylor expansion of $J_k(\textbf{p})$ around the local solution $\textbf{p}^{(r)}$.
Similarly, the upper bound on EVE’s achievable rate is expressed as
\vspace{-0.01cm} \begin{equation}
{R}_k^{\text{EVE},\text{UB}} = S_k(\textbf{p}) - W_k(\textbf{p}),
\label{eq:UR_EVE_decomp}
\vspace{-0.01cm} \end{equation}
with $S_k(\textbf{p}) \triangleq \log_2\big(p_k g_k^+ + \sum_{\substack{i \in \mathcal{K}, i \neq k}} p_i g_i^- + \sigma^2\big)$, and $W_k(\textbf{p}) \triangleq \log_2\big(\sum_{\substack{i \in \mathcal{K}, i \neq k}} p_i g_i^- + \sigma^2\big)$.
The non-convexity of ${R}_k^{\text{EVE},\text{UB}}$ stems from $S_k(\textbf{p})$. Accordingly, we approximate $S_k(\textbf{p})$ via its first-order Taylor expansion:
\small
\vspace{-0.1cm} \begin{equation}
\hat{S}_k^{(r)}(\textbf{p}) \triangleq S_k\big(\textbf{p}^{(r)}\big) + \frac{(p_k - p_k^{(r)})\, g_k^+ + \sum_{\substack{i \in \mathcal{K} \\ i \neq k}} (p_i - p_i^{(r)}) g_i^-}{\ln 2 \big(p_k^{(r)} g_k^+ + \sum_{\substack{i \in \mathcal{K} \\ i \neq k}} p_i^{(r)} g_i^- + \sigma^2\big)}.
\label{eq:S_taylor}
\vspace{-0.1cm} \end{equation}
\normalsize
By replacing $J_k(\textbf{p})$ with $\hat{J}_k^{(r)}(\textbf{p})$ and $S_k(\textbf{p})$ with $\hat{S}_k^{(r)}(\textbf{p})$ in ($\mathcal{SP}1$), we obtain the following convex approximation at the $r$-th iteration of the SCA:
\begin{subequations}
\vspace{-0.1cm} \begin{align}
(\mathcal{SP}1.r):\quad &\underset{\mathbf{p}, \mathbf{t}^\text{Th}} \Min \quad \sum_{k \in \mathcal{K}} t^\text{Th}_k \label{eq:SP1r_obj} \\[1mm]
\text{s.t.} \quad & 
\begin{aligned}[t]
&I(\textbf{p}) - \hat{J}_k^{(r)}(\textbf{p}) -\hat{S}_k^{(r)}(\textbf{p}) + \\
&W_k(\textbf{p}) \geq \frac{d_k}{B(t_k^{\text{Th}}-T_k^{\text{Off}})}, \quad \forall k \in \mathcal{K}_\text{Off},
\end{aligned} \label{eq:SP1r_cons}\\
& \eqref{eq:power_constraint}. \nonumber
\vspace{-0.1cm} \end{align}
\end{subequations}
This convex problem can then be solved optimally using convex solvers such as CVX.

\vspace{-0.1cm}
\subsection{Sub-problem 2: Computing Capacity Allocation}
\label{Subproblem2} \vspace{-.1cm} 
Given fixed $\alpha_k$ and $p_k$, the sub-problem of computing capacity allocation derived from ($\mathcal{P}$) can be simplified as:
\begin{subequations}
\vspace{-0.1cm} \begin{align}
(\mathcal{SP}2):\quad \underset{\mathbf{f}} \Min \quad & \sum_{k \in \mathcal{K}} T_k^{\text{Off}} \label{eq:SP2_obj} \\[1mm]
\text{s.t.} \quad & \eqref{eq:P_comp_cap}. \nonumber
\vspace{-0.2cm} \end{align}
\end{subequations}
This strictly convex problem admits a unique optimal solution via the Karush–Kuhn–Tucker (KKT) conditions. The Lagrangian is $\mathcal{L} = \sum_k \frac{d_k c_k}{f_k} + \lambda(\sum_k f_k - f^{\text{Tot}})$, and the stationarity condition gives $f_k^* = \sqrt{d_k c_k/\lambda}$. Substituting this into the constraint \eqref{eq:P_comp_cap} yields the optimal allocation:
\vspace{-0.1cm} \begin{equation} \label{eq:SP2_opt}
f_k^* = \frac{\sqrt{d_k c_k}}{\sum\limits_{j \in \mathcal{K}} \sqrt{d_j c_j}} f^{\text{Tot}}.
\vspace{-0.1cm} \end{equation}

\subsection {Sub-problem 3: Offloading Decision Optimization} \vspace{-.1cm} \label{Subproblem3}
with fixed values of $f_k$ and $p_k$, problem ($\mathcal{P}$) reduces to optimizing only the binary offloading decision vector:
\begin{subequations}
\vspace{-0.2cm} \begin{align}
(\mathcal{SP}3):\quad \underset{\boldsymbol{\alpha}} \Min \quad & \sum_{k \in \mathcal{K}} T_k^{\text{Tot}} \label{eq:SP3_obj} \\[1mm]
\text{s.t.} \quad & \eqref{eq:P_comp_cap}, \eqref{eq:P_binary}. \nonumber
\end{align}
\end{subequations}
Due to the NP-hardness of $(\mathcal{SP}3)$, we propose a greedy heuristic-based approach to efficiently approximate the optimal solution. First, we rewrite the objective in \eqref{eq:SP3_obj} as $\sum_{k} \left( T_k^{\text{Loc}} + \alpha_k \Delta T_k \right)$, where $\Delta T_k = T_k^{\text{Tra}} + T_k^{\text{Off}} - T_k^{\text{Loc}}$ represents the latency gain from offloading compared to local processing. Next, we sort tasks in ascending order of $\Delta T_k$, and set $\alpha_k=1$ for those tasks with the most negative $\Delta T_k$ values, until the resource constraint \eqref{eq:P_comp_cap} is satisfied. 

\begin{figure}[!t]
\begin{algorithm}[H]
\caption{Proposed AO-SCA Algorithm for Problem ($\mathcal{P}$)}
\small
\label{alg:AO-SCA}
\begin{algorithmic}[1]
\STATE \textbf{Initialize:} feasible $\mathbf{p}^{(0)}$, $\mathbf{f}^{(0)}$, and $\boldsymbol{\alpha}^{(0)}$, iteration index $r\leftarrow0$.
\REPEAT
    \STATE \textbf{Step 1:} Solve ($\mathcal{SP}1.r$) using the current $\mathbf{f}^{(r)}$ and $\boldsymbol{\alpha}^{(r)}$ to update $\mathbf{p}^{(r+1)}$.
    \STATE \textbf{Step 2:} Update $\mathbf{f}^{(r+1)}$ via~\eqref{eq:SP2_opt}, using $\mathbf{p}^{(r+1)}$ and $\boldsymbol{\alpha}^{(r)}$.
    \STATE \textbf{Step 3:} Solve relaxed ($\mathcal{SP}3$) and apply greedy rounding to update $\boldsymbol{\alpha}^{(r+1)}$.
    \STATE $r \leftarrow r+1$
\UNTIL{the change in objective value falls below the threshold $\delta_\text{AO}$.}\vspace{-.05cm}
\end{algorithmic}
\end{algorithm}
\vspace{-1.1cm}
\end{figure}
\normalsize

\vspace{-.2cm}
\subsection {Overall Algorithm} \vspace{-.1cm} 
\label{Overall}
Algorithm~\ref{alg:AO-SCA} summarizes the proposed AO-SCA method, which alternately solves subproblems ($\mathcal{SP}1$)–($\mathcal{SP}3$) to address the joint optimization in ($\mathcal{P}$).
The algorithm follows an AO scheme in which each subproblem is solved optimally at every iteration. Since the overall objective is non-increasing and bounded below by zero, convergence is guaranteed. 
We now analyze the computational complexity. 
Sub-problem 1 is solved iteratively via SCA, with each convex approximation ($\mathcal{SP}1.r$) handled by an interior-point solver. This incurs a complexity of $\mathcal{O}_{\mathcal{SP}1}(R_{\text{SCA}} K^{3.5})$, with $R_{\text{SCA}}$ denoting the number of SCA iterations. 
Sub-problem 2 admits a closed-form solution via KKT conditions, leading to a linear complexity of $\mathcal{O}_{\mathcal{SP}2}(K)$.
Sub-problem 3 applies a greedy rounding procedure after sorting the tasks by $\Delta T_k$, yielding $\mathcal{O}_{\mathcal{SP}3}(K \log K)$.
Let $R_{\text{AO}}$ be the number of AO iterations; then, the overall complexity is $\mathcal{O}(R_{\text{AO}}(R_{\text{SCA}}K^{3.5}+K \log K))$.


\section{Simulation Results} \vspace{-.1cm} \label{Simulation_Results}
We evaluate the proposed AO-SCA algorithm under a 3GPP small-cell channel model with Rayleigh fading and path loss $L(d)=30.6+36.7\log_{10}(d)+\mu$ (dB), where $d$ (in meters) is the transmission distance and $\mu \sim \mathcal{N}(0, 8 \,\text{dB})$ represents the log-normal shadow fading component \cite{3GPP}. We set $K=10$, $\sigma^2 = -110$ dBm, $f_t = 2.45$ GHz (ARMv8), $f_0 = 168$ MHz (ARM Cortex-M4), and $B = 500$ MHz, unless otherwise specified. The PQES and EVE are located at $(0\,\text{m}, 0\,\text{m})$ and $(50\,\text{m}, 0\,\text{m})$, respectively, and the RCDs are randomly placed within a 50\,m radius centered at the PQES. The maximum channel estimation error $\epsilon$ is set to $10\%$ of the corresponding path loss, and the AO-SCA stopping threshold is $\delta_{\text{AO}}=10^{-6}$. All results are averaged over 5000 Monte Carlo runs. 
Each RCD's data size, $d_k$, follows a uniform distribution $d_k \sim \text{U}[10,50]$ KB. The corresponding CPU cycles per bit, $c_k$, are randomly assigned from Table~\ref{tab:comparison}, reflecting each RCD may employ a different PQC algorithm and require a distinct security level \cite{Kannwischer2019, C_k_Kyber, C_k_RSA_WolfSSLBenchmarks}.
We compare the performance of our proposed AO-SCA algorithm (\emph{Proposed approach}) against three baselines (BL), along with a traditional no-EVE setup that serves as an upper-bound (UB) reference:
\\\textit{BL~I - Constant transmission power (CTP):} All RCDs transmit at the maximum power, $p_k = p^\text{MAX},\ \forall k\in\mathcal{K}$; 
Computing capacity allocation and offloading decisions are still optimized using Algorithm~\ref{alg:AO-SCA}.
\\\textit{BL~II - Uniform computing capacity (UCC):} The PQES divides its computing resources uniformly among all associated users; The optimal RCDs’ transmit power and the binary decision variables are jointly optimized using Algorithm~\ref{alg:AO-SCA}. 
\\\textit{BL~III - Full local computing (FLC):} Each RCD processes cryptographic tasks locally, i.e., $\alpha_k=0,\ \forall k\in\mathcal{K}$.
\\\textit{UB - No EVE:} Setting $g_k = 0,\ \forall k \in \mathcal{K}$ in problem ($\mathcal{P}$) yields an upper bound on performance without considering EVE.
\begin{figure*}[htb]
  \centering
  \begin{minipage}{.32\textwidth}
    \includegraphics[width=2.2in]{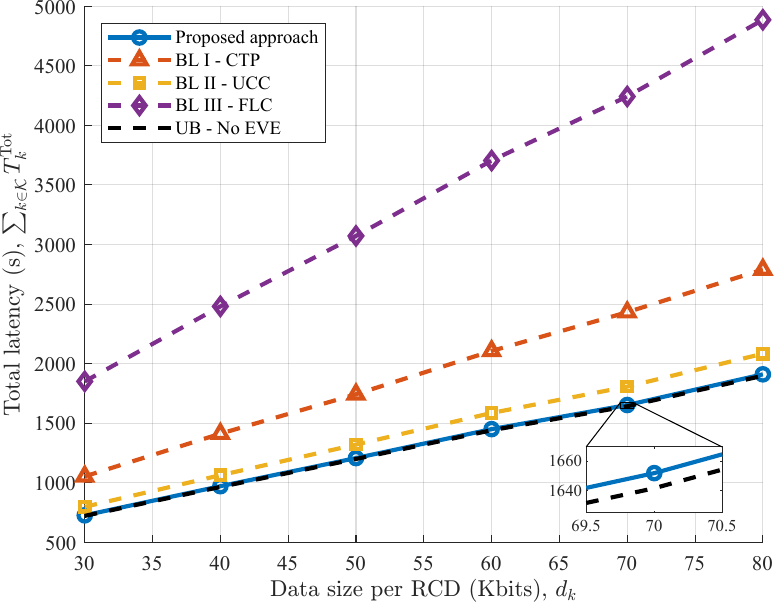}\vspace{-0.3cm}
    \caption{System total latency versus transmit data size per RCD, $d_k$.} 
    \label{fig2}
  \end{minipage}\hfill
  \begin{minipage}{.32\textwidth}
    \includegraphics[width=2.2in]{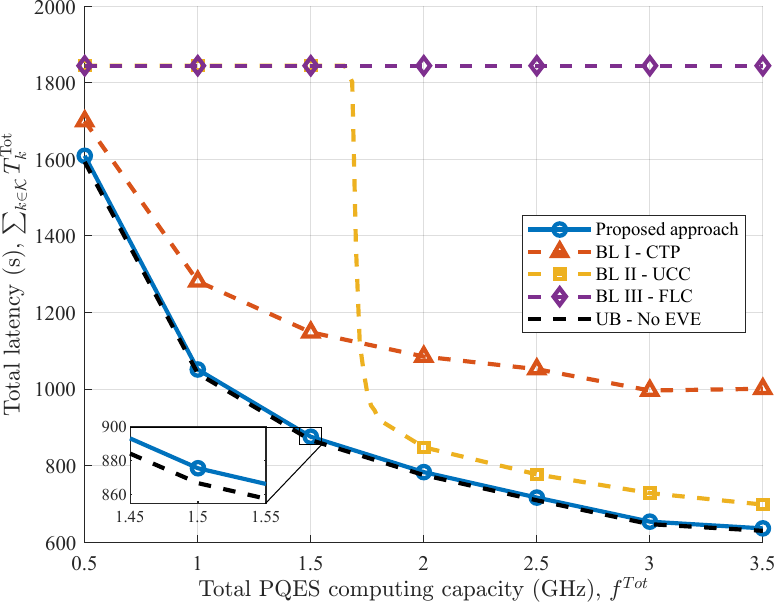}\vspace{-0.3cm}
    \caption{System total latency versus the PQES computing capacity, $f^\text{Tot}$.}
    \label{fig3}
  \end{minipage}\hfill
  \begin{minipage}{.32\textwidth}
    \includegraphics[width=2.2in]{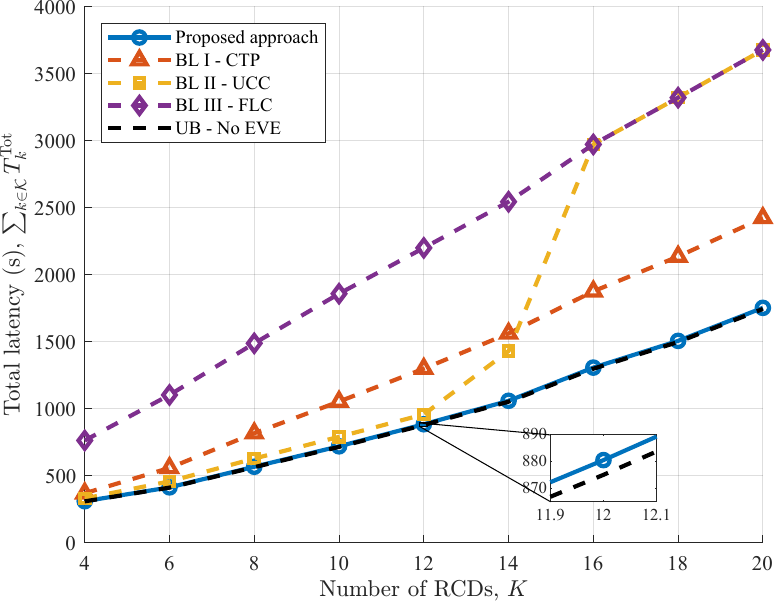}\vspace{-0.3cm}
    \caption{System total latency versus versus the number of RCDs, $K$.}
    \label{fig4}
  \end{minipage}
\vspace{-.5cm}
\end{figure*}

Fig.~\ref{fig2} illustrates the impact of the data size per RCD on the total latency for the proposed approach and the baseline schemes. As the data size $d_k$ increases, all schemes exhibit a monotonic increase in total latency due to higher communication and computation demands. The proposed approach consistently achieves the lowest latency across all data sizes, demonstrating its effectiveness in balancing offloading decisions, transmit power control, and computing resource allocation. 
Baseline I (CTP), which uses a fixed transmit power, performs approximately 60\% worse than the proposed method due to its lack of adaptability in transmission control. Baseline II (UCC), which applies uniform computing capacity allocation (while still optimizing the RCDs' transmit power and binary decision variables via Algorithm~\ref{alg:AO-SCA}), delivers moderate performance, about 10\% worse than the proposed scheme, but still falls short. Baseline III (FLC), which assumes full local computing, incurs the highest latency, emphasizing the importance of edge offloading in PQEC scenarios involving resource-constrained devices. 
Even when compared with the upper bound (i.e., the No EVE scenario), the proposed method exhibits only a minor performance gap. This demonstrates that our PLS-assisted offloading framework achieves robust edge-domain security at virtually no additional performance cost and underscores the effectiveness of the co-designed computation offloading and PLS strategy in mitigating eavesdropping overhead.

Next, Fig.~\ref{fig3} presents total latency as a function of the PQES's computing capacity $f^\text{Tot}$.
As expected, increasing $f^\text{Tot}$ significantly reduces latency for offloading schemes, with the proposed approach outperforming all baselines at every capacity level. Although baseline I (CTP) sees reduced latency as $f^\text{Tot}$ increases, its performance is still markedly inferior due to fixed power settings. Baseline II (UCC) exhibits a sharp latency drop around 1.7 GHz as it benefits from the additional computing resources; however, its uniform resource allocation still results in suboptimal performance. These results underscore the performance advantage of the proposed approach, which is particularly evident in lower computing capacities, where intelligent resource allocation becomes crucial.
Meanwhile, Baseline III (FLC),  which relies solely on local computing, shows no latency improvement with increases in $f^\text{Tot}$, highlighting its disadvantage without edge computing support.
Notably, the proposed approach remains close to upper-bound across varying $f^\text{Tot}$, demonstrating that the PLS-assisted strategy effectively mitigates the impact of EVE and preserves near-optimal performance.

Finally, Fig.~\ref{fig4} illustrates how the total latency scales with the number of RCDs $K$. As the network grows, all schemes experience increased latency due to higher communication and computation demands.  Nonetheless, the proposed approach consistently achieves the lowest latency, with its advantage becoming more pronounced at larger values of $k$,  where dynamic joint allocation of transmit power and edge computing resources is critical. Baseline I (CTP) and II (UCC) both display rising trends, but their fixed strategies in power or computing allocation lead to suboptimal performance under high device counts. In particular, Baseline II exhibits a sudden latency spike after $K=14$, highlighting bottlenecks resulting from uniform capacity allocation; as the number of devices grows, dynamic allocation becomes essential. Baseline III (FLC), relying solely on local computing, performs worst, demonstrating poor scalability and high latency, underscoring the limitations of local processing in large networks.
Moreover, the proposed approach remains close to the upper bound even as $K$ grows, underscoring its ability to effectively handle adversarial threats through PLS-assisted offloading.
\vspace{-.1cm}
\section{Conclusion} \vspace{-.1cm} \label{Conclusion}
In this paper, we proposed a novel co-design strategy that integrates PLS techniques and computation offloading to address the challenges of implementing PQC in IoT environments. By jointly optimizing device transmit power, PQES computing capacity allocation, and offloading decisions, our AO-SCA algorithm tackles the non-convex, NP-hard formulation through alternating optimization and successive convex approximation. Numerical results confirm that the proposed method substantially reduces latency and improves security compared to baseline schemes, even under channel uncertainty. Furthermore, our proposed approach achieves performance nearly identical to the upper-bound (No EVE) scenario, confirming that robust edge-domain security can be attained with our framework at a negligible additional cost.

\vspace{-.1cm}
\bibliographystyle{IEEEtran}
\bibliography{IEEEabrv,Bibliography}


\vfill

\end{document}